\documentclass[8.5pt,twoside,twocolumn]{article}
\usepackage[super,sort&compress,comma]{natbib}
\usepackage{mhchem}
\usepackage{times,mathptmx}
\usepackage{secsty}
\usepackage{balance}

\usepackage{graphicx} 
\usepackage{lastpage}
\usepackage[format=plain,justification=raggedright,singlelinecheck=false,
font=small,labelfont=bf,labelsep=space]{caption}
\usepackage{fancyhdr}
\begin{document}

\thispagestyle{plain}
\fancypagestyle{plain}{
\fancyhead[L]{\includegraphics[height=8pt]{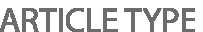}}
\fancyhead[C]{\hspace{-1cm}\includegraphics[height=20pt]{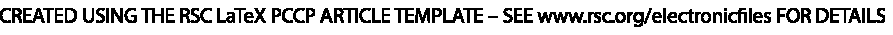}}
\fancyhead[R]{\includegraphics[height=10pt]{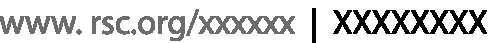}\vspace{-0.2cm}}
\renewcommand{\headrulewidth}{1pt}}
\renewcommand{\thefootnote}{\fnsymbol{footnote}}
\renewcommand\footnoterule{\vspace*{1pt}%
\hrule width 3.4in height 0.4pt \vspace*{5pt}}
\setcounter{secnumdepth}{5}

\makeatletter
\def\subsubsection{\@startsection{subsubsection}{3}{10pt}{-1.25ex plus -
1ex minus -.1ex}{0ex plus 0ex}{\normalsize\bf}}
\def\paragraph{\@startsection{paragraph}{4}{10pt}{-1.25ex plus -1ex minus
-.1ex}{0ex plus 0ex}{\normalsize\textit}}
\renewcommand\@biblabel[1]{#1}
\renewcommand\@makefntext[1]%
{\noindent\makebox[0pt][r]{\@thefnmark\,}#1}
\makeatother
\renewcommand{\figurename}{\small{Fig.}~}
\sectionfont{\large}
\subsectionfont{\normalsize}

\fancyfoot{}
\fancyfoot[LO,RE]{\vspace{-7pt}\includegraphics[height=9pt]{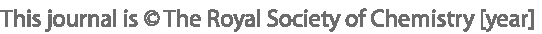}}
\fancyfoot[CO]{\vspace{-
7.2pt}\hspace{12.2cm}\includegraphics{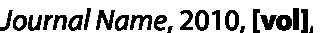}}
\fancyfoot[CE]{\vspace{-7.5pt}\hspace{-
13.5cm}\includegraphics{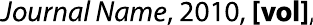}}
\fancyfoot[RO]{\footnotesize{\sffamily{1--\pageref{LastPage} ~\textbar
\hspace{2pt}\thepage}}}
\fancyfoot[LE]{\footnotesize{\sffamily{\thepage~\textbar\hspace{3.45cm}
1--\pageref{LastPage}}}}
\fancyhead{}
\renewcommand{\headrulewidth}{1pt}
\renewcommand{\footrulewidth}{1pt}
\setlength{\arrayrulewidth}{1pt}
\setlength{\columnsep}{6.5mm}
\setlength\bibsep{1pt}

\twocolumn[
  \begin{@twocolumnfalse}
\noindent\LARGE{\textbf{Self-assembly of spherical interpolyelectrolyte
complexes from oppositely charged polymers}}
\vspace{0.6cm}

\noindent\large{\textbf{Vladimir A. Baulin,$^{\ast}$\textit{$^{a,b}$} and
Emmanuel Trizac\textit{$^{c}$}}}\vspace{0.5cm}

\noindent\textit{\small{\textbf{Received Xth XXXXXXXXXX 20XX, Accepted
Xth XXXXXXXXX 20XX\newline
First published on the web Xth XXXXXXXXXX 200X}}}

\noindent \textbf{\small{DOI: 10.1039/b000000x}}
\vspace{0.6cm}

\noindent \normalsize{The formation of inter-polyelectrolyte
complexes from the association of oppositely charged polymers in
an electrolyte is studied. The charged polymers are linear
oppositely charged polyelectrolytes, with possibly a neutral
block. This leads to complexes with a charged core, and a more
dilute corona of dangling chains, or of loops (flower-like
structure). The equilibrium aggregation number of the complexes
(number of polycations $m_+$ and polyanions $m_-$) is determined
by minimizing the relevant free energy functional, the Coulombic
contribution of which is worked out within Poisson-Boltzmann
theory. The complexes can be viewed as colloids that are permeable
to micro-ionic species, including salt. We find that the
complexation process can be highly specific, giving rise to very
localized size distribution in composition space ($m_+,m_-$).}
\vspace{0.5cm}
 \end{@twocolumnfalse}
  ]



\footnotetext{\textit{$^{a}$ICREA, Passeig Lluis Companys, 23 Barcelona,
Spain E-mail: vladimir.baulin@urv.cat}}
\footnotetext{\textit{$^{b}$Departament d'Enginyeria Quimica, Universitat
Rovira i Virgili, Av.
dels Paisos Catalans, 26 Tarragona, Spain}}
\footnotetext{\textit{$^{c}$Universit\'{e} Paris-Sud, Laboratoire de
Physique Th\'{e}orique et Mod\`{e}les Statistiques, UMR CNRS 8626, 91405
Orsay, France }}


\section{Introduction}

Electrostatic interactions are instrumental in determining the
structure and function of living organisms, biopolymers and drug
delivery systems. Charged macromolecules can self-assemble and
aggregate into compact intermolecular complexes. This ability of
oppositely charged polymers to form finite size complexes
determines their biological function, which for example is
important in gene transfection and compactization of DNA
\cite{KabanovDNA,Hatton,Gohy}, that provide promising alternatives
to viral vectors \cite{Balaban}. Such macromolecular systems,
where electrostatic forces are usually stronger than van der Waals or
hydrogen bonds, exhibit rich behavior and structural variability.
The structures formed by opposite charges are
usually more stable than neutral block copolymers micelles
dissociating upon dilution or slight change in the external
conditions. The concept of stabilization of intermolecular
complexes by interaction of oppositely charged polymers is
realized in inter-polyelectrolyte or polyion complexes (PIC) and
polyion complex micelles (PIC micelles) that can be used for drug
delivery \cite{Kabanov,Kabanov3,HaradaRev}. High stability of PICs
opens the possibility to use them as functional devices where the
responsiveness to external stimuli can be connected with a
function, \textit{e.g.} recognition at the molecular level
\cite{Harada}, pH-sensitive switching devices \cite{Frechet} or
drug delivery carriers transporting charged objects through the
cell membrane \cite{Kabanov2,HaradaRev}.

\begin{figure}
\begin{center}
\includegraphics[width=6cm]{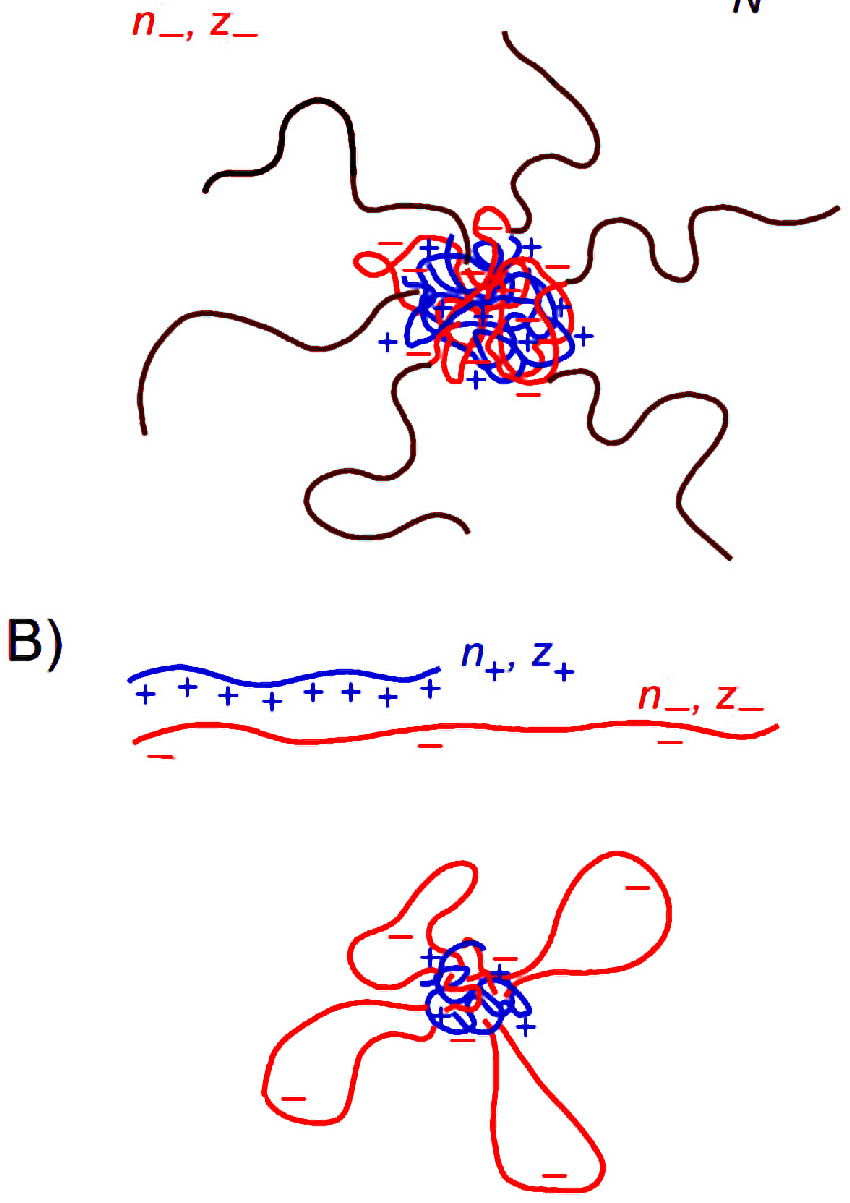}
\end{center}
\caption{Interpolyelectrolyte complexes formed by A) a linear
polyelectrolyte (blue) and a diblock copolymer composed of an oppositely
charged block (red) and a neutral block (black);
upon assembling, these chains form the complex sketched, where
the corona is made up of the neutral blocks;
B) two linear oppositely
charged polyelectrolytes (blue and red) with large asymmetry in the
distances between charges ($n_{+}$ and $n_{-}$). The segments with non-
compensated charges
form a charged corona of loops.}
\label{Fig:Polymic}
\end{figure}

Polyion complexes have enhanced ability to undergo structural changes
subject
to external conditions, compared to neutral block copolymer assemblies.
In
addition to the response in change of temperature and solvent quality
\cite%
{Liu}, the structure of the charged complexes can be very sensitive to
changes in salt concentration
\cite{Muller2,Leclercq,Vasilevskaya,Kabanov4},
pH \cite{Kabanov4,Frechet,Liu2,Gohy}, charge ratio \cite{Stuart,Schwarz},
addition of ions \cite{Schwarz2}, or mixing ratio \cite{Hatton,Kataoka}.
Tuning
the molecular architecture and global properties of PICs would allow for
precise control of their
functional properties. Understanding the physics and fundamental features
of
the self-assembly of PICs is thus a challenging task.
Oppositely charged polymers in symmetric solutions can precipitate into
uniform macroscopic phase of polymers and ions \cite{Kudlay,Kudlay2}. The
physics of aggregated chains of opposite charge in precipitate is somehow
similar to polyampholites, polymers containing both positive and negative
charges dispersed along the chain \cite{Joanny}. However, if the
distribution of charges along the chain is not random \cite{Bae} or one
of
the charged polymers is a diblock copolymer with a neutral block \cite%
{Harada,Gohy}, oppositely charged polymers can form finite size
complexes composed of a dense polyelectrolyte core and a swollen
corona which protects the cores from aggregation by steric
repulsion.


In this paper, we explore electrostatic properties and equilibrium
structures of spherical complexes formed by oppositely charged
polymers. The stability of finite size aggregates results from the
balance of the electrostatic attraction between opposite charges
in the core of the complexes and the steric repulsion of backbone
segments forming a corona around the core. The description of such
complexes is similar to polyelectrolyte micellization
\cite{Kramarenko1,Kramarenko2,Zhulina,Zhulina2}, combined with
thermodynamics of aggregation in bidisperse solution \cite%
{Castelnovo,Sens,Marques,Baulin}. The steric repulsion in the
corona should be strong enough to stabilize the complexes of
finite size. This is possible when the hydrophilic blocks forming
the corona of the complexes are long enough to oppose the
electrostatic attraction in the core. In addition, the bare charge
of the complexes can be screened by the ions of salt and
counterions, thus changing the electrostatic forces and affecting
the equilibrium properties of the complexes. The interplay of
those effects will be studied considering two geometries, as
sketched in Fig. \ref{Fig:Polymic}:
\begin{itemize}
\item (case A) the building blocks are a linear uniformly
charged polyelectrolyte and a diblock copolymer composed of a
neutral block and a charged block of opposite charge. The charged
blocks aggregate with the linear polyelectrolytes to form a complex
with a core surrounded by a corona of neutral segments.
This case is that of a \textquotedblleft hairy\textquotedblright\
and neutral surrounding outside the charged core.
\item (case B) the building blocks are
two linear polyelectrolytes of opposite charge
with a large asymmetry of the distances between the charges. The
core of such complexes is composed of charged blocks of both signs,
and is surrounded by the corona of loops of the segments between
the charges. The corona can be neutral (segments between
neighboring charges along the chain) or slightly charged (tails or
longer segments between distant charges). Compared to case A, the
dangling \textquotedblleft hair\textquotedblright\ are replaced
by loops, that may bear an electric charge.
\end{itemize}

The paper is organized as follows. In section \ref{sec:A}, we first
consider
case A, with a charged core decorated by neutral dangling hair. The
electrostatics of the complexes is taken into account through the full
Poisson-Boltzmann (PB) equation which is solved numerically and compared
with the analytical expression of the linearized Debye-H\"uckel (DH)
equation.
Particular attention will be paid to the counter-ion uptake,
where a significant quantity of charge can be ``trapped'' inside the
core,
thereby reducing the electric field created outside the complex. A second
mechanism for charge reduction is ascribable to the non-linearity of
Poisson-Boltzmann framework: non-linear screening effectively modifies
the
total core charge, leading in general to a reduced effective (or
renormalized) quantity seen from a large distance\cite{Trizac02}.
At this level of description, the dangling hair are not taken into
account.
The more complex situation where the charged core is surrounded by
charged loops
(case B) will be addressed in section \ref{sec:B}. In turn, these results
will be used in section \ref{sec:complex} to discuss the complexation
behaviour of oppositely charged polymers. Conclusions will finally be
drawn
in section \ref{sec:concl}. An appendix summarizes the main notations
employed.


\section{Charged core surrounded by neutral corona (case A)}
\label{sec:A}

\subsection{The model and its three relevant charges}

The simplest structure of a thermodynamically stable polyion
complexes of finite size is a spherical core, containing all bare
charges, surrounded by a neutral corona (Figure
\ref{Fig:Polymic}A). Such a complex can be formed,
for example, by diblock copolymers containing neutral blocks \cite%
{HaradaRev,Gohy,Harada,Kabanov3,Kabanov1,Bae,Stuart,Wu}. The
electrostatic interactions between oppositely charged
polyelectrolytes drive the formation of a dense core which is
stabilized by the steric repulsion of neutral blocks forming
swollen corona around the core. Even in such simple geometry, it
is possible to tune the structure of the complex. Its size can be
controlled by the lengths of the blocks, the density of charges,
pH, the charge asymmetry, the salt concentration and solution
properties.

Assuming that the linear polyelectrolyte is positively charged
while the blocks of the diblock copolymer bear a negative charge,
a linear polyelectrolyte is described by the number of charges on
the chain $z_{+}$, and the distance between the charges $n_{+}$
while the block copolymer is described by the number of charges
$z_{-}$, the distance between the charges $n_{-}$, and the length
of a neutral block, $N$. Hence, the length of the polyelectrolyte
chain is $n_{+}z_{+}$ and the total length of a block copolymer is
$n_{-}z_{-}+N$. Here, the lengths are expressed in units of a Kuhn
length (assumed common to both cationic and anionic chains).

If all polyelectrolyte charges of both signs are buried in the
core and the neutral blocks form the corona, the total "bare"
charge of the core formed by $m_{+}$ linear polyelectrolytes and
$m_{-}$ block copolymer chains is
\begin{equation}
Z_{1}=z_{+}m_{+}-z_{-}m_{-}  \label{Z1}
\end{equation}%
We will assume in the subsequent analysis that this charge is
uniformly spread over the globule of radius $R_{c}$, and therefore
occupies a volume $4\pi R_{c}^{3}/3$. The counterions and the salt
ions can penetrate into the core of radius $R_{c}$, and thus,
screen the bare charge of the polymers. The resulting charge of
the "dressed" core, $Z_{2}$, is the charge of the core screened by
small ions; assuming spherical symmetry, we have
\begin{equation}
Z_{2}=4\pi \int_{0}^{R_{c}}r^{2}dr\rho (r)  \label{Z2}
\end{equation}%
where the total charge density $\rho (r)$ reads (in units of the
elementary charge $q$)
\begin{equation}
\rho (r)=\frac{Z_{1}}{4\pi R_{c}^{3}/3}\,H(R_{c}-r)\,+\,c_{\infty }e^{-
\beta
q\varphi (r)}\,-\,c_{\infty }e^{\beta q\varphi (r)}.  \label{rho}
\end{equation}%
In the above relation, $\beta =1/kT$ is the inverse temperature,
$\varphi (r)$ is the electrostatic potential and $H(R_{c}-r)$
denotes the Heaviside step function, equal to $1$ inside and $0$
outside the core. The first term on the right hand side of Eq.
(\ref{rho}) stems from the polymeric matrix, that contributes to
the charge density as a spherical uniform background. We assume
here that the system is in osmotic equilibrium with a salt
reservoir with equal densities $c_{\infty }$ of labile cations and
anions; the canonical situation, where the salt density in the
system would be \textit{a priori} prescribed, is amenable to a
very similar treatment as the one presented here. From the
reservoir ionic concentration, we define the Debye length
$1/\kappa $ through $\kappa ^{2}=8\pi \ell _{B}c_{\infty }$, where
$\ell _{B}=\beta q^{2}/\epsilon $ is the Bjerrum length and
$\epsilon $ is the solvent dielectric permittivity. In Eq.
(\ref{rho}), the last two terms are for the labile micro-ions
concentration. The corresponding exponential relation between the
density profiles and the local electrostatic potential is typical
of the PB (Poisson-Boltzmann) approximation \cite{Levin} that will
be adopted in the remainder. Within such a mean-field
simplification, the electrostatic problem at hand is the
following:
\begin{equation}
\left\{
\begin{array}{l}
\nabla ^{2}\varphi (r)=-\frac{4\pi }{\epsilon }q\rho (r) \\
\displaystyle\left. \frac{d\varphi }{dr}\right\vert _{r=0}=0 \\
\displaystyle\left. \frac{d\varphi }{dr}\right\vert _{r\rightarrow \infty
}=0%
\end{array}
\right.   \label{PB}
\end{equation}

The charge of the "dressed" core $Z_{2}$ should be smaller than
the "bare" charge $Z_{1}$ \cite{Trizac4}, Eq. (\ref{Z1}), because
of counter-ion penetration inside the globule. However, the latter
charge may be large enough to trigger significant non-linear
screening effects, that translate into an effective (or
renormalized \cite{Trizac02,Trizac3}) charge $Z_{3}$ that can be
much smaller than $Z_{2}$. To be more precise, in the weakly
coupled limit where $Z_{2}\rightarrow 0$ (where one also has
$Z_1\rightarrow 0$), it is possible to solve analytically Eq.
(\ref{PB}), since it reduces to $\nabla ^{2}\varphi =\kappa
^{2}\varphi $ for $r>R_c$. The resulting DH (Debye-H\"{u}ckel)
potential reads, for $r>R_c$
\begin{equation}
\varphi_{DH} (r)\, = \,Z_{1}\,\Theta (\kappa R_{c})\,\frac{e^{-\kappa
r}}{%
\kappa r},  \label{phi}
\end{equation}%
where $\Theta$ is a salt-dependent geometric prefactor; the
complete solution will be provided below in section \ref{ssec:A2}.
To define the renormalized charge $Z_{3}$, it is sufficient to
note that beyond the linear Debye-H\"{u}ckel regime, for an
arbitrary charge $Z_{2}$, Eq. (\ref{PB}) again takes the form
$\nabla ^{2}\varphi \simeq \kappa ^{2}\varphi $, but at large
distances $r$ where $\varphi $ becomes small. We consequently
have, within the non- linear PB framework:
\begin{equation}
\varphi (r)\,\sim \,Z_{3} \, \Theta (\kappa R_{c})  \,
\frac{e^{-\kappa r}}{\kappa r}
\qquad \hbox{for}\qquad  r\rightarrow \infty .
\label{eq:ff}
\end{equation}
By construction, $Z_{3}\simeq Z_{1}$ in the Debye-H\"{u}ckel
regime, while $Z_{3}\ll Z_{1}$ upon increasing $Z_{1}$.

Introducing the dimensionless electrostatic potential $u(r)=\beta
q\varphi (r)$ and dimensionless distance $x=r/R_{c}$, Eq.
(\ref{PB}) in spherical polar coordinates is written in a
dimensionless form
\begin{equation}
\left\{
\begin{array}{l}
\displaystyle u^{\prime \prime }(x)+\frac{2}{x}u^{\prime }(x)=-
3\widetilde{Z}
_{1}H(1-x)+(\kappa R_{c})^{2}\sinh (u(x)) \\
\displaystyle u^{\prime }(0)=0 \\
u^{\prime }(\infty )=0
\end{array}
\right.  \label{PBdim}
\end{equation}
There are then two dimensionless governing parameters,
$\widetilde{Z}_{1}=Z_{1}\ell_{B}/R_{c}$ and $\kappa R_{c}$. The
solution of this nonlinear equation gives the charge density and
the distribution of small ions around the complex.

\subsection{Two limiting cases: weak charges and salt-free situation}
\label{ssec:A2}

Equation (\ref{PBdim}) can be solved analytically in the DH
approximation when the electrostatic potential is small, $u(x) \ll
1$. In this case, the charge density $\rho (x)$ can be linearized,
$e^{\pm u(x)}\approx 1\pm u(x)$, and the solution can be written
in the form
\begin{equation}
u_{DH}(\kappa r)=\left\{
\begin{array}{l}
\vartheta \left[ 1-\left( 1+\kappa R_{c}\right) e^{-\kappa
R_{c}}\frac{\sinh
(\kappa r)}{\kappa r}\right] , \qquad r<R_{c} \\
\vartheta \left[ \kappa R_{c}\cosh (\kappa R_{c})-\sinh (\kappa
R_{c})\right]
\frac{e^{-\kappa r}}{\kappa r}, \qquad r>R_{c}%
\end{array}%
\right.  \label{uDH}
\end{equation}%
where $\vartheta =3\widetilde{Z}_{1}/(\kappa R_{c})^{2}$. As a
consequence, the geometrical constant that enters into the
electrostatic potential at large distances (\ref{phi}) is
\begin{equation}
\Theta (\kappa R_c)=\frac{3}{(\kappa R_{c})^{2}}\left[ \kappa R_{c}\cosh
(\kappa R_{c})-\sinh (\kappa R_{c})\right].
\end{equation}

Our results for a charged polyelectrolyte complex in the presence
of salt can be compared with the salt free regime \cite{Raphael}.
In this case, it is essential to enclose the complete system in a
confining boundary, otherwise the counter-ions \textquotedblleft
evaporate\textquotedblright\ --their energy loss upon leaving the
globule vicinity is out-beaten by the entropy gain of exploring a
large volume-- and the problem becomes trivial. We therefore
define $R_{WS}$, the Wigner-Seitz radius\cite{Trizac3} of a large
sphere containing the system. We note that the ratio $\eta
=(R_{c}/R_{WS})^{3}$ defines the volume fraction of globules in
our system. In the present case where counterions only are
present, the density of charges is
\begin{equation}
\rho (r)=\frac{3Z_{1}}{4\pi R_{c}^{3}}\,H(R_{c}-r)-c_{0}e^{\beta q\varphi
(r)},
\end{equation}
where $c_{0}$ is a normalization parameter to ensure total
electroneutrality (it does not have any physical significance as
such, unless a particular \textquotedblleft
gauge\textquotedblright\ or reference has been chosen for the
potential). A possible choice among others is $4\pi
c_{0}R_{WS}^{3}/3=Z_{1}$. The corresponding dimensionless PB
equation reads
\begin{equation}
\left\{
\begin{array}{c}
u^{\prime \prime }(x)+\frac{2}{x}u^{\prime }(x)=-3\widetilde{Z}_{1}\left[
H(1-x)-\eta e^{u(x)}\right] \\
u^{\prime }(0)=0 \\
u^{\prime }(R_{WS}/R_{c})=0%
\end{array}%
\right.  \label{PBws}
\end{equation}%
with re-scaled distance $x=r/R_{c}$. The dimensionless charge
$\widetilde{Z}_{1}=Z_{1}\ell _{B}/R_{c}$ and $\eta $ are here
independent control parameters.

\begin{figure*}[th]
\begin{center}
\includegraphics[width=17cm]{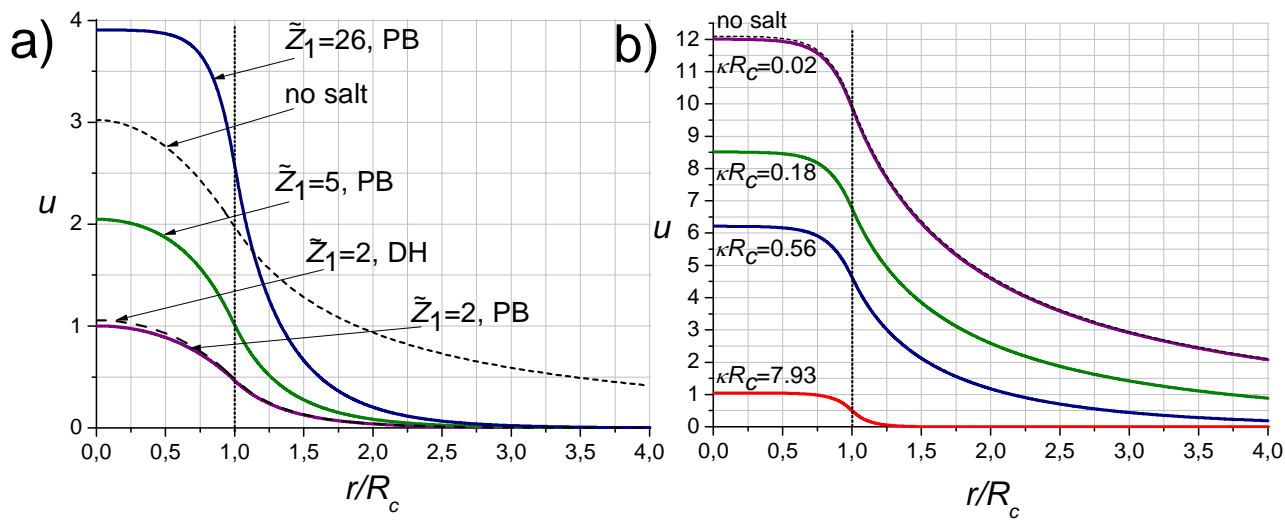}
\end{center}
\caption{a) {Dimensionless electrostatic potential $u$ as a
function of the charge of the core }$\widetilde{Z}_{1}$, for a
fixed salt concentration corresponding to $\protect\kappa
R_{c}=1.77$. b) {Electrostatic potential $u$ as a function of salt
concentration. The charge of the core is fixed,
$\widetilde{Z}_{1}=26$.}{ The dashed line is the solution of Eq.
(\protect\ref{PBws}) with no salt, solid lines are the solutions
of the non-linear PB\ equation
(\protect\ref{PBdim}%
), }b{lack dashed line is the solution of the DH equation
(\protect\ref{uDH}%
),} and the value of the packing fraction to solve the salt-free
problem is $\protect\eta = (R_c/R_{WS})^3 = 0.000125$.}
\label{FiguZ}
\end{figure*}

\subsection{Results}

The solutions of the PB equation (\ref{PBdim}) for different
dimensionless bare charge $\widetilde{Z}_{1}=Z_{1}\ell_B/R_{c}$
are presented in Figure \ref{FiguZ}a). The comparison with DH
approximation (\ref{uDH}) shows as expected that this
approximation is valid for weakly charged objects. Such a
comparison is a test for the numerical procedure used to solve
Eqs. (\ref{PB}), and of course, strong deviations are observed
between DH and PB solutions for $\widetilde Z_1$ larger than a few
units. It is noteworthy in Fig. \ref{FiguZ} that the electric
potential inside the core tends to a plateau when $Z_1$ is high
enough. The corresponding labile ion local charge indeed tends to
compensate for the background charge, resulting in a vanishing
total local charge density. This requirement implies that one has
\begin{equation}
u \to \hbox{arcsinh}[3 \widetilde Z_1/(\kappa R_{c})^{2}],
\label{eq:bound}
\end{equation}
which gives $u \to 3.91$ in Fig. \ref{FiguZ}a) for
$\widetilde{Z}_{1}=26$,
and likewise $u \to 8.48$ in Fig. \ref{FiguZ}b) for
$\kappa R_c = 0.18$, as can be seen.

\begin{figure*}[th]
\begin{center}
\includegraphics[width=17cm]{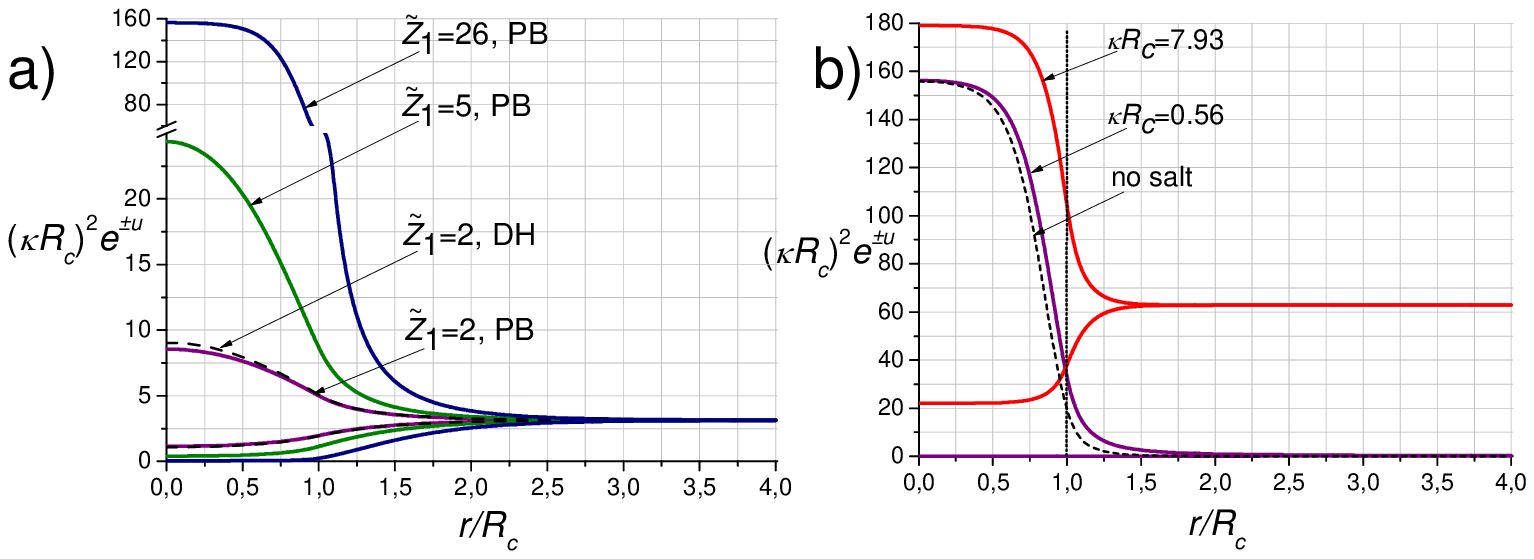}
\end{center}
\caption{{Reduced density (up to a factor of two)
of small ions around the core } $\protect\rho _{\pm
}=c_{\infty }e^{\mp u(x)}$ {a) for different charges of the core }$%
\widetilde{Z}_{1}${\ and }fixed salt concentration{\ }($\protect\kappa %
R_{c}=1.77$) {\ b) for different salt concentrations and }fixed charge of
the core{\ }($\widetilde{Z}_{1}=26$). For the no salt case,
$\protect\eta =0.000125$%
.}
\label{Figcion1}
\end{figure*}

The solution of the salt-free equation (\ref{PBws}) is shown by a
dotted line in Figure \ref{FiguZ}a). The absolute value of the
corresponding electric potential $u$ cannot be compared to its
counterpart found with salt, but the variations of $u$ can be. It
can be seen on the figure, panel a), that the amplitude of $u$ is
as expected larger without salt (for the same value of
$\widetilde{Z}_{1}$). This illustrates the weaker screening
without salt. In addition, panel b) shows that the small $\kappa
R_{c}$ limit coincides with the salt-free limit, as it should
(here, the salt-free solution has been shifted by the constant
required to have the same potential at $r=0$ as in the $\kappa R_c
=0.02$ case). The salt-free results reported here depend very
weakly only on packing fraction.

The solution of the PB equation provides also the distribution of
small ions around the core of the complex $\rho_{\pm}(r) =
c_{\infty}e^{\pm u(r)}$. It is shown for different charges of the
core $\widetilde{Z}_{1}$ in Figure \ref{Figcion1}a) and fixed salt
concentration $c_{\infty}$ in Figure \ref{Figcion1}b). As the
charge is increased the concentration of small ions of opposite
charge inside the core increases inducing stronger screening
effect. The concentration of ions of both signs in the core
increases with increase of the bulk concentration of salt. This is
shown in Figure \ref{Figcion1}b), where the bare charge of the
core $\widetilde{Z}_{1}$ is fixed, while the concentration of salt
in the solution is changed. If the salt concentration is low, the
redistribution of small ions around the core is almost due to
counterions and the concentration profile of small ions obtained
from Eq. (\ref{PBdim}) approaches the corresponding salt free
solution given by Eq. (\ref{PBws}). In addition, the increase of
salt concentration results in higher concentration and induces
stronger redistribution of the small ions around the core.

\begin{figure*}
\begin{center}
\includegraphics[width=8.25cm]{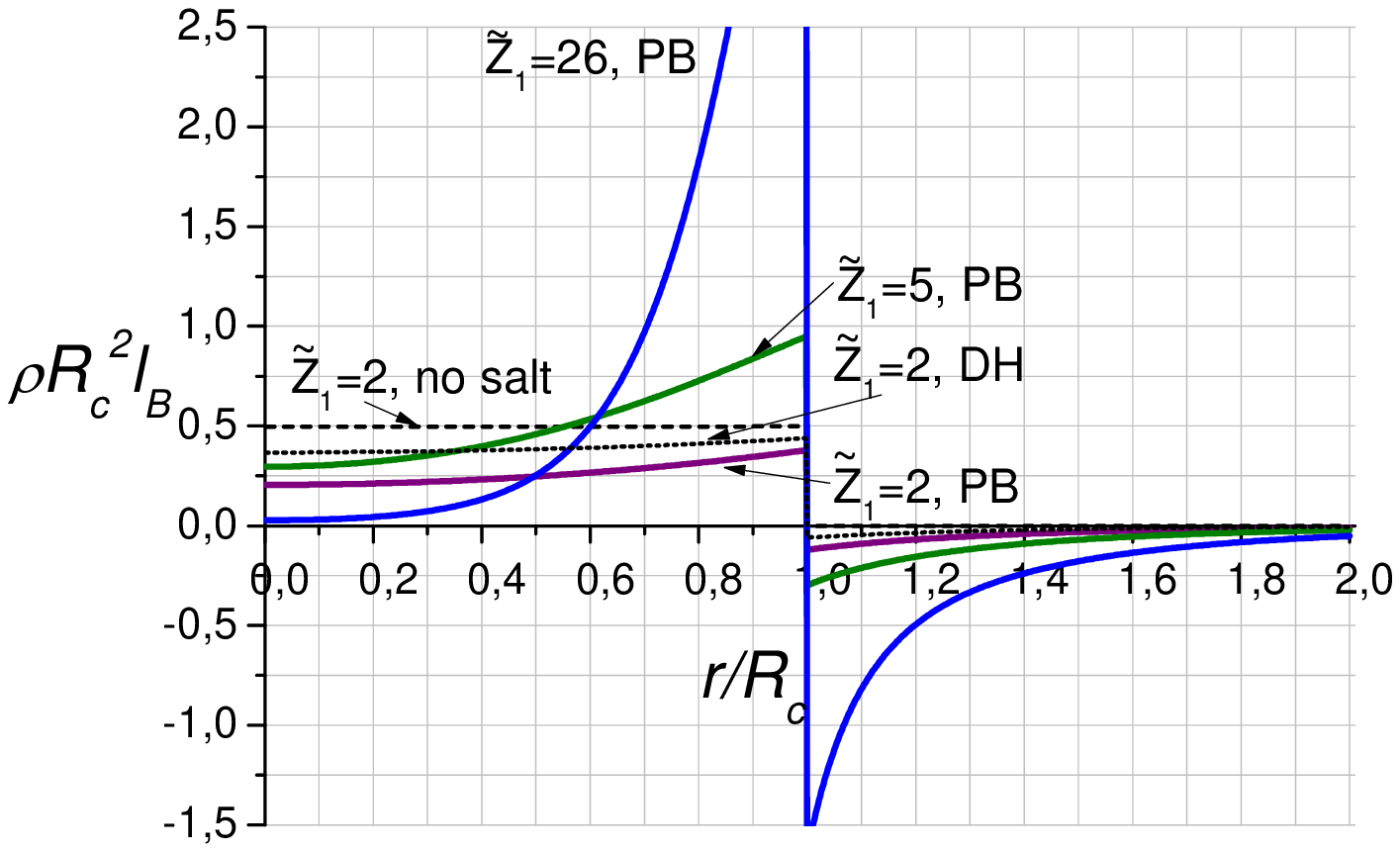}
\end{center}
\caption{{Charge density }$\protect\rho R_{c}^{2}\ell _{B}$ as a distance
from the core{\ for different charges of the core and fixed salt
concentration }$\protect\kappa R_{c}=1.77$. Here,
$\protect\eta =0.000125$ for the no-salt result.}
\label{Figrho}
\end{figure*}

The presence of small ions inside the complex impinges on the
radial distribution of charges $\rho (r)$. The background
\textquotedblleft core\textquotedblright\ contribution to this
quantity is a step function, while due to the penetration of
labile ions, $\rho (r)$ is small in the globule center, where
charge neutralization is most efficient, and increases upon
increasing $r$ (see Fig. \ref{Figrho}). In addition, the charge
density in the center of the core vanishes at high concentrations
but also for large enough $\widetilde{Z}_{1}$.

\begin{figure*}[th]
\begin{center}
\includegraphics[width=17cm]{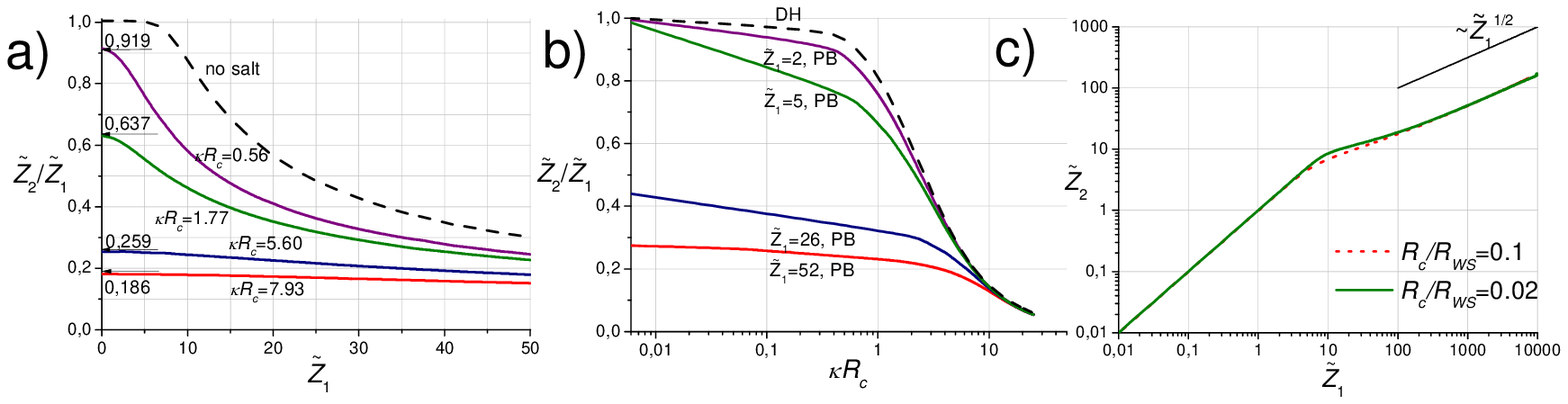}
\end{center}
\caption{Charge of the dressed core $\widetilde{Z}_{2}$ (uptake
charge) as defined in Eq. (\protect\ref{Z2}), as a function of a)
bare charge for different salt concentrations b) salt for
different values of bare charge $\widetilde{Z}_{1}$ c) bare charge
in the salt-free case, on a log-log scale. The arrows in panel a)
show the Debye-H\"uckel prediction, Eq. (\ref{limZ2Z1}), that
provides the correct limiting behaviour at small charges. Note
that in panel b), the Debye-H\"uckel prediction does not depend on
\smash{$\widetilde Z_1$}. The no salt solution in panel a)
corresponds to $\protect\eta =0.01$ (which means $R_c/R_{WS}\simeq
0.21$).} \label{FigZ1Z2Z}
\end{figure*}

The screening of the bare charge by small ions penetrating into
the core can be measured by the charge of the "dressed" core,
$Z_{2}$ introduced above in Eq. (\ref{Z2}). The analytical
expression for $Z_{2}$ can be obtained within the DH
approximation: $\rho (r)=3Z_{1}/(4\pi R_{c}^{3})-2c_{\infty
}u_{DH}(\kappa r)$, where $u_{DH}(\kappa r)$ is given by Eq.
(\ref{uDH}). Thus,
\begin{equation}
\rho (r)=\frac{3Z_{1}}{4\pi R_{c}^{3}}\left( 1+\kappa R_{c}\right)
e^{-\kappa R_{c}}\frac{\sinh (\kappa r)}{\kappa r},r<R_{c}
\end{equation}%
from which it follows that
\begin{equation}
\frac{Z_{2DH}}{Z_{1}}=3\frac{1+\kappa R_{c}}{(\kappa R_{c})^{3}}\left(
\kappa R_{c}\cosh (\kappa R_{c})-\sinh (\kappa R_{c})\right) e^{-\kappa
R_{c}}.  \label{limZ2Z1}
\end{equation}%
It can be checked that this relation is consistent with the
more familiar DH result for the potential of a spherical
colloid having bare charge $Z_2$ and radius $R_c$ \cite{Trizac02}:
\begin{equation}
\varphi(r) \, = \, Z_2 \, \frac{e^{\kappa R_c}}{1+\kappa R_c} \,
\frac{e^{-\kappa r}}{r},
\end{equation}
which imposes that
\begin{equation}
\frac{Z_{2DH}}{Z_{1}} \,=\, \Theta(\kappa R_c) \,
\frac{1+\kappa R_c}{\kappa R_c} \,
e^{-\kappa R_c}
\end{equation}

From Eq. (\ref{limZ2Z1}), we have that $Z_{2}\propto Z_{1}$ within
DH approximation, up to a salt-dependent prefactor. However, upon
increasing $Z_{1}$, non-linear effects become prevalent and
invalidate the DH approach, see Fig. \ref{FigZ1Z2Z}, obtained by
solving the non-linear PB theory. The corresponding slower than
linear increase of $Z_{2}$ with $Z_{1}$ is illustrated in Fig.
\ref{FigZ1Z2Z}a) and b), see also panel c) for the salt-free case.
Increasing salt concentration screens out the charge, i.e. it
leads to a decrease of $Z_2$, and flattens the curves in panel a),
where the DH prediction (\ref{limZ2Z1}) holds for low $\widetilde
Z_1$. In essence, increasing salt concentration ultimately leads
to the DH limit where $\widetilde{Z}_{2}/\widetilde{Z}_{1}$ is
$Z_{1}$ independent, as can be inferred from Eq. (\ref{uDH}),
where $u(0)$ is seen to decay with the increase of $\kappa $. A
similar conclusion is drawn from expression (\ref{eq:bound}): the
DH limit is reached in the high salt limit. This trend is clearly
seen in Fig. \ref{FigZ1Z2Z}b), where all curves tend to collapse
onto the DH behaviour for $\kappa R_c > 10$.
On the other hand, $\widetilde{Z}_{2}/\widetilde{Z}%
_{1}$ is a strongly nonlinear function for large
$\widetilde{Z}_{1}$ in the no salt case. More precisely, it has
been shown in Ref \citenum{Raphael} that in the strongly
non-linear salt-free regime, one has $Z_{2}\propto
Z_{1}^{1/2}$ (or equivalently $\widetilde{Z}_{2}\propto \widetilde{Z}%
_{1}^{1/2}$). This prediction is successfully put to the test in
Fig. \ref{FigZ1Z2Z}c), which also shows that a change in the
packing fraction $\eta = (R_c/R_{WS})^3$ does not affect the
features discussed.

\begin{figure*}[th]
\begin{center}
\includegraphics[width=17cm]{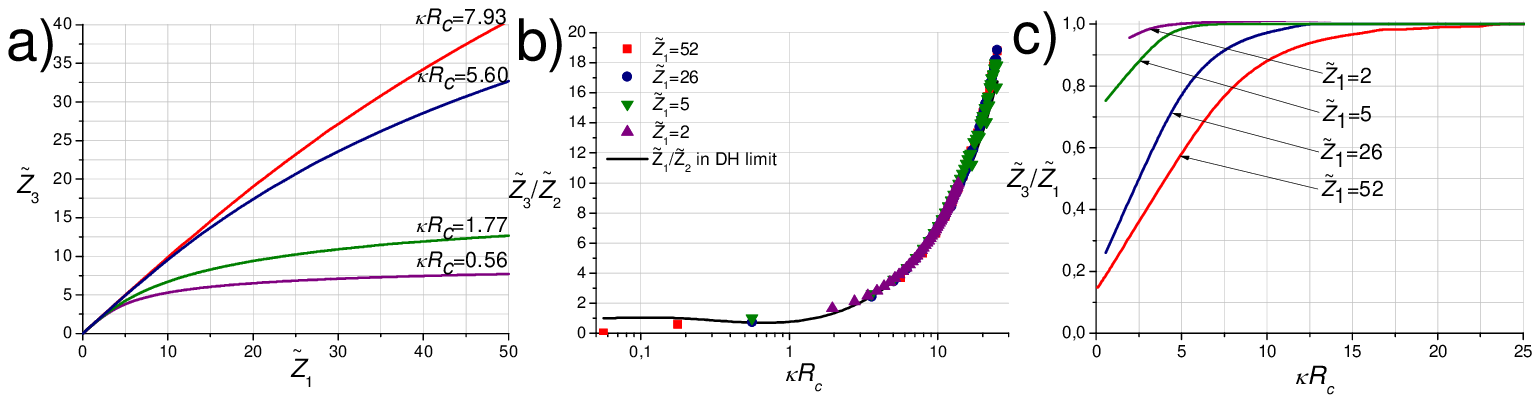}
\end{center}
\caption{a) Renormalized charge $\widetilde{Z}_{3}$ of a globule
as seen from large distances, as a function of the bare charge
$\widetilde{Z}_{1}$ for different salt concentrations. b) and c)
show either $Z_3/Z_2$ or $Z_3/Z_1$ as a function of salt
$\protect\kappa R_{c}$, for different bare charges. In panel b),
the continuous curve is for the ratio $Z_1/Z_2$ found within DH
approximation, which is thus the inverse of Eq. (\ref{limZ2Z1}).}
\label{FigZ3Z1Z}
\end{figure*}


Finally, the behavior of the charged globule at large distances is
encoded in the effective charge $Z_{3}$, defined in Eq.
(\ref{eq:ff}), and therefore extracted from the far-field of the
numerical solution to the non-linear equation (\ref{PBdim}). Such
a quantity would rule the interactions between two distant
globules. The corresponding plots of $\widetilde{Z}_{3}$ are shown
in Figure \ref{FigZ3Z1Z}a) as a function of the globule charge for
fixed salt concentration, and as a function of salt density for
fixed background charge in Fig. \ref{FigZ3Z1Z}b) and c). As is
invariably the case in such mean-field approaches, the effective,
or renormalized, charge increases upon increasing the bare charge
\cite{Trizac02,Trizac3}. It also increases with salt concentration
\cite{Trizac02} and as imposed by the very definition of $Z_{3}$,
we find that $Z_{3}/Z_{1}\rightarrow 1$ in the DH limit (enforced
either from considering low $\widetilde{Z}_{1}$ or large $\kappa
R_{c}$). In addition, Fig. \ref{FigZ3Z1Z}b) shows that the ratio
$Z_3/Z_2 = \widetilde Z_3/\widetilde Z_2$ is independent of bare
charge $Z_1$, except at very small salt concentrations. This
reflects the fact that even for large $Z_1$, counterion uptake is
such that $Z_2$ is significantly reduced, and such that the
colloid included internal salt ions can be treated by linearized
mean-field theory. Indeed, it can be seen in Fig. \ref{FigZ3Z1Z}b)
that $\widetilde Z_3/\widetilde Z_2$ is close to its DH
counterpart, given by $\widetilde Z_1/\widetilde Z_{2DH}$, see Eq.
(\ref{limZ2Z1}).

\section{Charged core surrounded by charged corona (case B)}
\label{sec:B}

Thermodynamically stable polyelectrolyte complexes can also be
formed by the complexation of two linear polyelectrolytes of
opposite charge with a large asymmetry of the distances between
the charges $\Delta =n_{-}/n_{+}\gg 1$, which can form flower-like
structures \cite{Bae}. The core of such complexes with a partially
compensated charges is surrounded by a corona of long loops of a
polymer with a longer distance between the charges (Fig.
\ref{Fig:Polymic}-B). The loops of size $n_{-}$, are neutral, but
some larger loops and the tails can be charged.

Thus, we can generalize the discussion of the previous chapter to
the case where the charged core is surrounded by a charged corona.
We assume spherical symmetry in the distribution of the charges
around the core, i.e. the charge in the corona depends only on the
distance from the center of the core $r$.
Since the charges in the loops and tails are attached to the core
by polymer chains, the electrostatic interaction of those charges
with the core is balanced by a weak entropic force due to polymer
chain extension. If the electrostatic force is not very strong, a
polymer chain carrying the charge can be envisioned as a Gaussian
coil and the probability of radial distribution of charges is
$P(r)\sim \exp \left[ -\frac{3}{2na^{2}}(R_{c}-r)^{2}\right] $,
where $n$ is the length of the polymer chain in the corona and $a$
is the Kuhn segment length. This approximation is valid for small
charges that do not perturb significantly the statistics of the
chains. The resulting density of charges is the sum of three
terms: the bare charge of the core, the charge of the counterions
plus salt molecules, and the charge of the corona:
\begin{eqnarray}
\rho (r) &=&\frac{3Z_{1}}{4\pi R_{c}^{3}c_{\infty }}\,H(R_{c}-
r)+c_{\infty
}e^{-\beta q\varphi (r)}-c_{\infty }e^{\beta q\varphi (r)}  \notag \\
&& + \rho _{c}P(r)e^{-\alpha \beta q\varphi (r)}H(r-R_{c})
\end{eqnarray}%
where $\rho _{c}$, positive or negative, is a parameter
controlling the total charge of the corona and $\alpha =\pm $ is
the sign of this charge ($+$ when $\rho_c>0$ and $-$ when
$\rho_c<0$). It was assumed here that all loops have the same
length $n$ and that ions in the corona are monovalent. This
expression for $\rho(r)$ leads to a Poisson equation similar to
Eq. (\ref{PBdim}).

\begin{figure*}[th]
\begin{center}
\includegraphics[width=17cm]{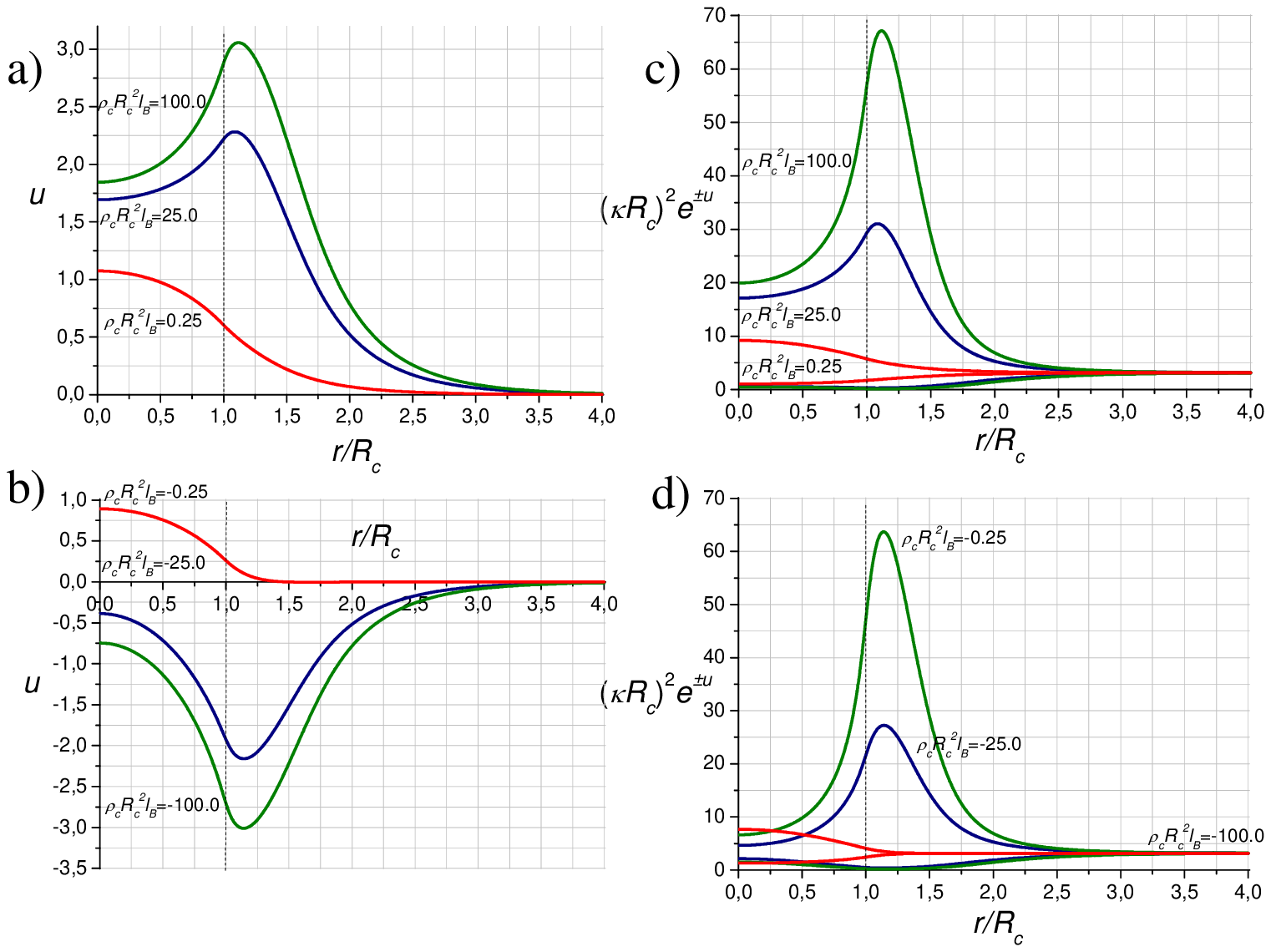}
\end{center}
\caption{Electrostatic potential $u(r)$ of the aggregate with a
charged core $\widetilde{Z}_{1}=2$, and a) a positively charged
and b) a negatively charged corona. Distribution of small ions
$(\protect\kappa R_{c})^{2}e^{\pm u}$ around the positively
charged core, $\widetilde{Z}_{1}=2$, surrounded by c) positively
charged and d) negative charged corona. Here $\protect\kappa
R_{c}=1.77$, $R_c= 5\, a$ and $n = 5$.} \label{Fig:u_rhoplus}
\end{figure*}

Solution of Poisson's equation gives the radial distribution of
the potential which is shown in Figure \ref{Fig:u_rhoplus}a) and
\ref{Fig:u_rhoplus}b) for different charges of the corona. The
charged corona influences the distribution of small ions around
the core, a quantity that is displayed in Fig.
\ref{Fig:u_rhoplus}c) and \ref{Fig:u_rhoplus}d). Weakly charged
coronas clearly do not modify the monotonous decrease of $u$ with
distance that was observed in Fig. \ref{FiguZ}, but this is no
longer the case when $\rho_c$ is increased. Indeed, a point where
$u$ reaches an extremum [maximum in panel a) and minimum in panel
b)] can be observed. From Gauss theorem, this coincides with the
point where the total integrated charge over a sphere having the
corresponding radius vanishes. The physical phenomenon occurring
in panel a) where the charges in the corona are of the same sign
as the bare core (assumed positive), is that the positive corona
induces a migration of negative micro-ions inside the core and its
vicinity, that change the sign of the uptake charge $Z_2$, which
is now negative. Adding the corona charge to $Z_2$, though, leads
to a positive charge. Hence the charge inversion evidenced by the
potential extremum. The density peak of negative micro-ions is
clearly visible in panel c). On the other hand, when the bare core
and the corona bear charges of opposite signs [panels b) and d)],
a conjugate mechanism takes place: small labile cations are
``sucked'' inside by the core, which leads to an integrated charge
in a running sphere that is positive for small spheres, and
becomes negative once it includes the corona. In all cases, the
mechanism can be viewed as a corona induced local charge
inversion.

\begin{figure*}[th]
\begin{center}
\includegraphics[width=17cm]{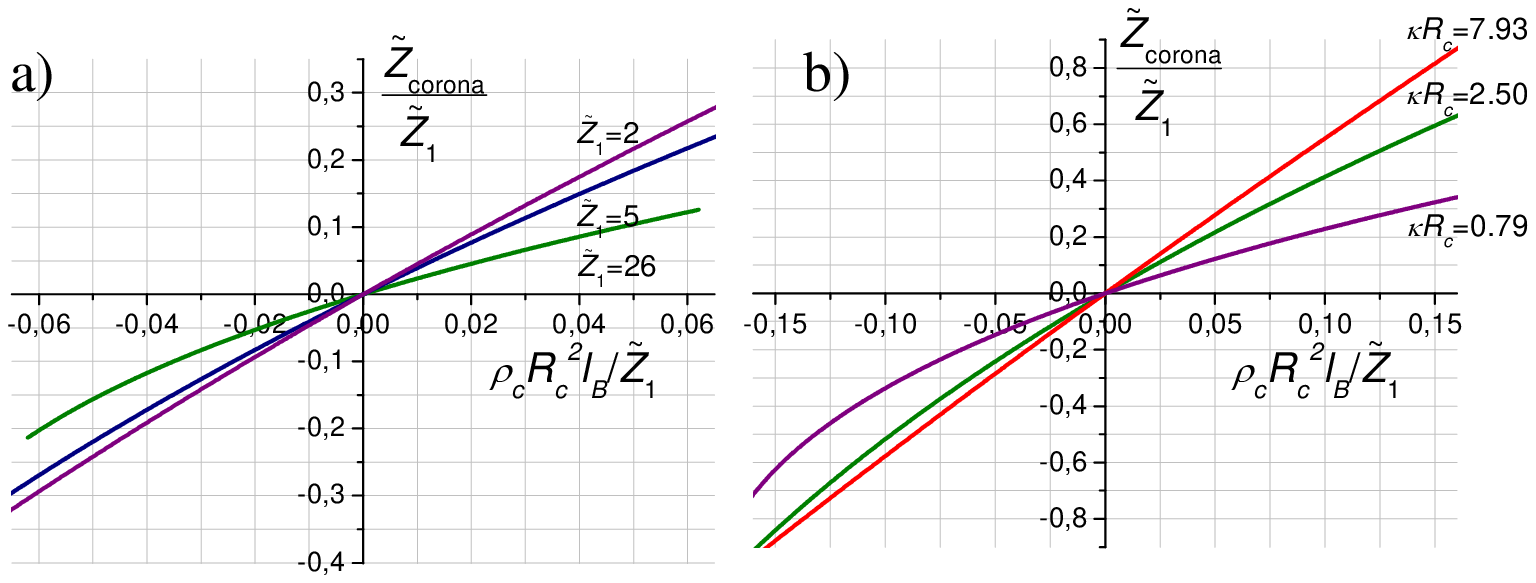}
\end{center}
\caption{Total charge of the corona
$\widetilde{Z}_{corona}/\widetilde{Z}%
_{1} $ as a function of $\protect\rho _{c}R_{c}^{2}\ell _{B}$, a)
the salt concentration is fixed, $\protect\kappa R_{c}=2.5$, b)
the bare charge of the core is fixed, $\widetilde{Z}_{1}=2$.}
\label{Fig:Zcorona}
\end{figure*}

We do not repeat the full analysis of the difference between bare,
uptake, and effective charge in the present case, but we show how
the total charge of the corona, $Z_{corona}=$
$\int_{R_{c}}^{\infty }r^{2}dr\alpha \rho _{c}P(r)e^{-\alpha \beta
q\varphi (r)}H(r-R_{c})$, depends on the parameter $\rho_{c} $ in
Fig. \ref{Fig:Zcorona}.

\section{Complexation of oppositely charged polymers}
\label{sec:complex}

The previous sections, devoted to the electrostatics of
polyelectrolyte complexes, have left aside the energetical
aspects, to which we turn our attention hereafter. Once the total
free energy of a given complex is known, it becomes possible to
study the equilibrium behaviour, in particular the size
distribution, of an initial ``soup'' of individual polycations and
polyanions.

\subsection{The total free energy and equilibrium complex size
distribution}

The size of the thermodynamically stable complexes of oppositely
charged polymers is determined by the interplay of the steric
repulsion of the chains in the corona and electrostatic attraction
in the core. Thus, the formation of stable aggregates requires
either long neutral blocks at least in one of the polyelectrolyte
or a large asymmetry in the distances between the charges along
the chain, e.g. $\Delta =n_{-}/n_{+}\gg 1$. If the corona is
composed of neutral blocks, the blocks should be long enough to
stabilize the attraction in the core, if the corona is composed of
loops between the charges, the segment $n_{-}$ should be long and
flexible enough to form a loop in the corona in the micelle.

Consider then a spherical polyion complex made up of two
polyelectrolytes of opposite charge. Each complex is defined by
the number of polycations, $m_{+}$, and the number of polyanions,
$m_{-}$. If we assume dense packing of the monomers in the core,
the radius of the core $R_{c}$ can be expressed in terms of the
number of chains $m_{+}$ and $m_{-}$ as $R_{c}=a\left[
\frac{3}{%
4\pi }(N_{+}m_{+}+N_{-}m_{-})\right] ^{\frac{1}{3}}$, where $a$ is
the Kuhn segment, $N_{\pm }=n_{\pm }q_{\pm }$ are the lengths of
the charged blocks (in units of $a$). The bare charge of the core
$Z_{1}$  is then also expressed in terms of $m_{+}$ and $m_{-}$,
see Eq. (\ref{Z1}).

The distribution function of the polyion complexes
$c_{m_{+},m_{-}}$ is the number concentration of the aggregates
with given aggregation numbers $m_{+}$ and $m_{-}$. The total free
energy of the solution of polyelectrolytes of opposite charge,
their counterions and salt molecules is
\begin{equation}
\frac{F}{VkT}=\sum_{m_{+},m_{-}=0}^{\infty }\left( c_{m_{+},m_{-}} \ln
\left[%
c_{m_{+},m_{-}}v\right] - c_{m_{+},m_{-}}
+c_{m_{+},m_{-}}F_{m_{+},m_{-}}\right)
\label{eq:free}
\end{equation}
where $V$ is the volume of the system, $F_{m_{+},m_{-}}$ is the
free energy of the complex expressed in units of $kT$, $v$ is a
molecular volume associated with the de Broglie length.
Minimization of this free energy\cite{Castelnovo,Sens} with
respect to $c_{m_{+},m_{-}}$ along with two conservation of mass
conditions, fixing the total concentrations of polyanions, $\phi
_{-}$, and polycations, $\phi _{+}$,
\begin{equation}
\phi _{\pm }=\sum_{m_{\pm }=0}^{\infty }m_{\pm }c_{m_{+},m_{-}}
\label{eq:phis}
\end{equation}
gives the equilibrium distribution of the aggregates by their
size\cite{Baulin}

\begin{equation}
vc_{m_{+},m_{-}}=(vc_{1,0})^{m_{+}}(vc_{0,1})^{m-}
\exp \left[ -\left(F_{m_{+},m_{-}}-m_{+}F_{1,0}-m_{-}F_{0,1}\right)
\right]
\label{eq:cmp}
\end{equation}
The free energy of the complex $F_{m_{+},m_{-}}$
can be written as the sum of an electrostatic contribution,
and a term accounting for the steric repulsion of tails/loops in
the corona
\begin{equation}
F_{m_{+},m_{-}}=\Omega _{el}+F_{corona}.
\end{equation}%
These two contributions are detailed below.

\subsection{The electrostatic contribution}
\begin{figure*}[th]
\begin{center}
\includegraphics[width=17cm]{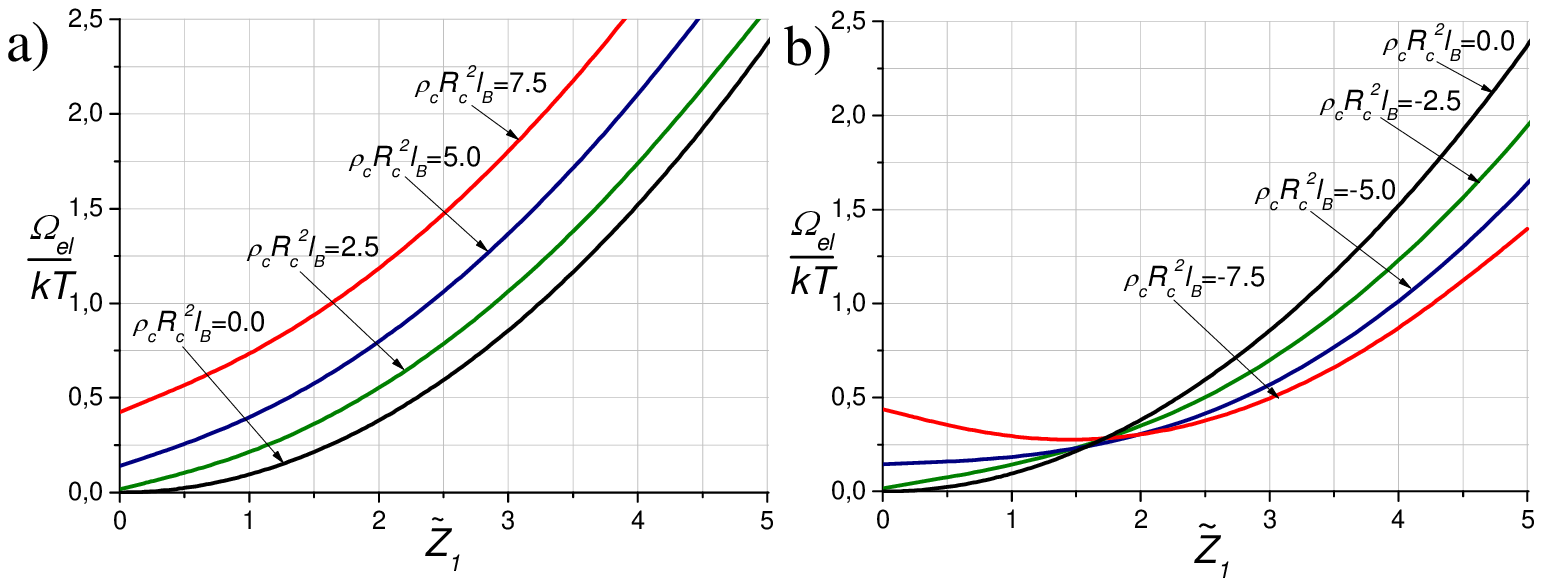}
\end{center}
\caption{Electrostatic excess energy $\Omega _{el}$ of a PIC with
a) positively charged corona and b) negatively charged corona, as
a function of the charge of the core, $\widetilde{Z}_{1}$, for a
fixed salt concentration, $\protect\kappa R_{c}=7.93$.}
\label{Fig:OmegaEl}
\end{figure*}

The electrostatic contribution $\Omega _{el}$ is related to the
semi-grand potential $\Omega _{el}^{\prime }$, relevant to discuss
the present situation which is canonical for the colloids
(polymers), and grand-canonical for the salt entities (in osmotic
equilibrium with a salt reservoir of density $c_{\infty }$). The
semi-grand potential accounts for electrostatic attraction between
polyelectrolytes and small ions in the system as well as the
entropic contribution of small ions around the complexes. We have
\cite{Trizac2}
\begin{equation}
\Omega _{el}^{\prime }=\int d\mathbf{r}\left\{ \frac{1}{2}q\rho
(\mathbf{r}%
)\varphi (\mathbf{r})+kT\sum\limits_{\alpha =\pm}\rho _{\alpha
}(\mathbf{r}%
)\left( \ln \left[ \frac{\rho _{\alpha }(\mathbf{r})}{c_{\infty }}\right]
-1\right) \right\}   \label{Fel}
\end{equation}%
where the first term is the electrostatic energy of the ionic
distribution, and the second term is the entropy associated with
the translational movements of small ions. We note that the
integral in Eq. (\ref{Fel}) diverges for large systems (as it
would also for neutral systems), so that we consider in the
following the excess semi-grand potential with respect to
reservoir
\begin{equation}
\Omega _{el}=\Omega _{el}^{\prime }-\Omega _{el}^{reservoir}=\Omega
_{el}^{\prime }+\int d\mathbf{r}2c_{\infty }
\end{equation}%
Thus, the excess potential $\Omega _{el}$ finally takes the form
\begin{eqnarray}
&&\Omega _{el}=\int d\mathbf{r}\left\{ \frac{1}{2}q\rho
(\mathbf{r})\varphi (%
\mathbf{r})+kT\sum\limits_{\alpha }\rho _{\alpha }(\mathbf{r})\ln
\frac{\rho
_{\alpha }(\mathbf{r})}{c_{\infty }} \right. \\ \nonumber
&&\left. -kT\sum\limits_{\alpha }\rho _{\alpha }(%
\mathbf{r})+2c_{\infty }\right\}   \label{Felg}
\end{eqnarray}%
For a spherical globule, this equation can be written in the
dimensionless form as
\begin{eqnarray}
&&\frac{\Omega _{el}}{kT}=4\pi R_{c}^{2}\ell _{B}\int_{0}^{\infty
}x^{2}dx\left\{ \frac{1}{2}\rho (x)u(x)\right. \\ \nonumber
&&\left. +\rho _{+}(x)\left[ -u(x)-1\right]+ \rho _{-}(x)\left[ u(x)-
1\right] +2c_{\infty }\right\}   \label{Feldim}
\end{eqnarray}%
where $x=r/R_{c}$ is the rescaled distance and we used the equality $%
\ln \left( \rho _{\pm }(x)/c_{\infty }\right) =\mp u(x)$. The
analysis of the free energy of the charged complexes suggests that
the complexes with charged corona of the same sign as the bare
core have larger free energy than their neutral counterparts, and
thus, are less favorable (Figure \ref{Fig:OmegaEl}a)). However, if
the charges of the core and the corona are opposite, the
electrostatic energy can be lower (Figure \ref{Fig:OmegaEl}b)).

The above applies for spherical globules, but leaves aside the
particular cases ($m_+=0$, $m_-=1$) and conversely ($m_+=1$,
$m_-=0$), where the object to be considered is no longer a
complex, but a polyanion or polycation respectively. We then need
to adapt the previous arguments to these cases of   an isolated
charged chain in a salt solution. The polyelectrolyte chain is
approximated as a cylinder of radius $a$ with uniform linear
charge density $\lambda _{\pm }\propto 1/n_{\pm }$, again treated
within Poisson- Boltzmann theory. Introducing dimensionless
distance $\widetilde{r}=\kappa r$, where $\kappa ^{2}=8\pi \ell
_{B}c_{\infty }$, the corresponding PB equation in cylindrical
coordinates yields, for an infinite cylinder
\begin{equation}
\left\{
\begin{array}{c}
\displaystyle\frac{1}{\widetilde{r}}\frac{d}{d\widetilde{r}}\left(
\widetilde{r}\frac{d}{d\widetilde{r}}\right) u=\sinh u \\
\displaystyle\left. \frac{du}{d\widetilde{r}}\right\vert _{\widetilde{r}%
=\kappa a}=\pm \frac{2\xi }{\kappa a} \\
u(\widetilde{r}\longrightarrow \infty )=0%
\end{array}%
\right.
\end{equation}%
Here $\xi =\ell _{B}\lambda _{\pm }$ is the so-called Manning
parameter (dimensionless line charge\cite{Tellez}). Once this
equation has been solved, the electrostatic contribution to
$F_{1,0}$ and $F_{0,1}$ for isolated chains of both signs follows
from a similar calculation as that of Eq. (\ref{Felg}):
\begin{eqnarray}
&&\frac{\Omega _{el,\pm }^{0}}{kTN_{\pm}}=\frac{1}{2}u(0)\xi
+\frac{1}{4}%
\int_{\kappa a}^{\infty }\widetilde{r}d\widetilde{r}u(\widetilde{r})\sinh
(u(%
\widetilde{r}))-\\ \nonumber
&&\frac{1}{2}\int_{\kappa a}^{\infty }\widetilde{r}d\widetilde{%
r}u(\widetilde{r})\left[ \cosh (u(\widetilde{r})-1\right]   \label{OmEl}
\end{eqnarray}%
which was calculated per chain length $N_{\pm}$ expressed in units of
$\ell _{B}$. In the following we assume that the Kuhn segment
length of the polymer $a$ is of the order of $\ell _{B}$. Upon
using the free energy of the infinite cylindrical macro-ion
configuration, we neglect end effects, the consideration of which
would be technically more involved.

\begin{figure*}[th]
\begin{center}
\includegraphics[width=17cm]{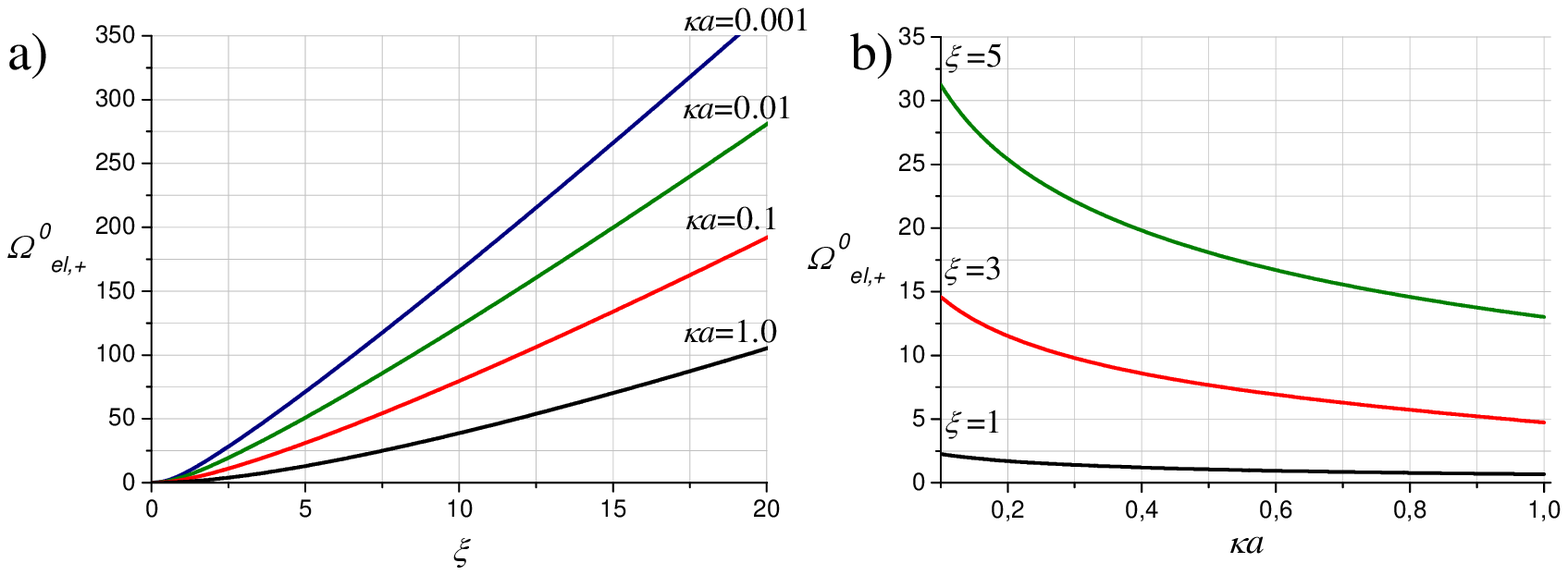}
\end{center}
\caption{Electrostatic energy $\Omega _{el,\pm }^{0}$ of the rod
(Eq. \protect\ref{OmEl}) for the length $\ell _{B}$ as a function
of a) the Manning parameter $\protect\xi $ (dimensionless linear
charge) and b) salt concentration $\protect\kappa a$.}
\label{Fig:OmElRod}
\end{figure*}

Isolated chains, corresponding to $m_+=0$, $m_-=1$ and $m_+=1$,
$m_-=0$ configurations are penalized by a large electrostatic
energy $\Omega _{el,+}^{0}$ (see Figure \ref{Fig:OmElRod}).
Indeed, these quantities bear a large self-term, notwithstanding
the solvation phenomenon, that manifests itself in the fact that
$\Omega _{el,+}^{0}$ decrease, for fixed charge $\xi $, upon
addition of salt (i.e. increase of $\kappa a$).

\subsection{Steric repulsion of loops and tails}
In the following, we consider a neutral and spherical
polyelectrolyte complex made up of the charged core formed by
oppositely charged polyelectrolytes and surrounded by the corona.
We consider long tails and large loops in the corona, thus, the
corona of the complex is then approximated by a star polymer (Fig.
\ref{Fig:Polymic}-A) or a flower structure (Fig.
\ref{Fig:Polymic}-B).

The electrostatic contribution (\ref{OmEl}) is balanced by the
steric repulsion between the tails or loops in the corona. If the
corona consists of long neutral blocks (star polymer, case A),
this contribution is approximated by the free energy of a star
polymer containing $m_{-}$ arms of length $N$, which is the length
of a neutral block. This approximation is valid when the core is
much smaller than the corona and the arms are long enough to use
the scaling expression\cite{Duplantier}
\begin{equation}
F_{corona}\sim -\ln N^{\sigma _{m_{-}}+m_{-}\sigma _{1}}  \label{Fstar}
\end{equation}%
In this expression, $\sigma _{i}$ are the universal exponents of the star
polymers and their numerical values are calculated in Ref. %
\citenum{Grassberger}.

If the corona consist of long neutral loops and tails (flower structure,
case B), the free energy contribution is similar to (\ref{Fstar}), but
the exponent is different,
\begin{equation}
F_{corona}\sim -\ln N^{\gamma _{c}-1}  \label{Fstar2}
\end{equation}%
This exponent is calculated as follows. If the loops are formed by
a single chain with $p$ stickers joined together, $\gamma
_{c}-1=\sigma _{2p}+2\sigma _{1}-(p-1)d\nu $, where the first term
is the contribution of the center with $2p$ vertices, the second
term is the contribution of the two tails and the last term is the
contribution of $p-1$ loops. Each loop contributes with the Flory
exponent $\nu $ in the dimension of the space $d$ and is known
numerically \cite{Rubin}. If the loops are formed by $m_{+}$
chains with $z_{+}$ stickers and $m_{-}$ chains with $z_{-}$
stickers and all stickers are condensed on the core, the exponent
is given by
\begin{eqnarray}
\gamma _{c}-1&=&\sigma _{2z_{+}m_{+}+2z_{-}m_{-}}+2\sigma
_{1}(m_{+}+m_{-})-\\ \nonumber
&&m_{+}(z_{+}-1)d\nu -m_{-}(z_{-}-1)d\nu
\end{eqnarray}%
Isolated non-aggregated polycations and polyanions are linear
polymers, thus their entropy contribution is $F_{corona}\sim -\ln
N^{2\sigma _{1}}$. Here we neglect the surface tension and
hydrophobic interactions between polycations and polyanions in the
core of the complexes, thereby assuming that the electrostatic
attraction of opposite charges is the leading contribution;
hydrophobic interactions may however be dominant for neutral
complexes.

\begin{figure*}[th]
\begin{center}
\includegraphics[width=17cm,clip]{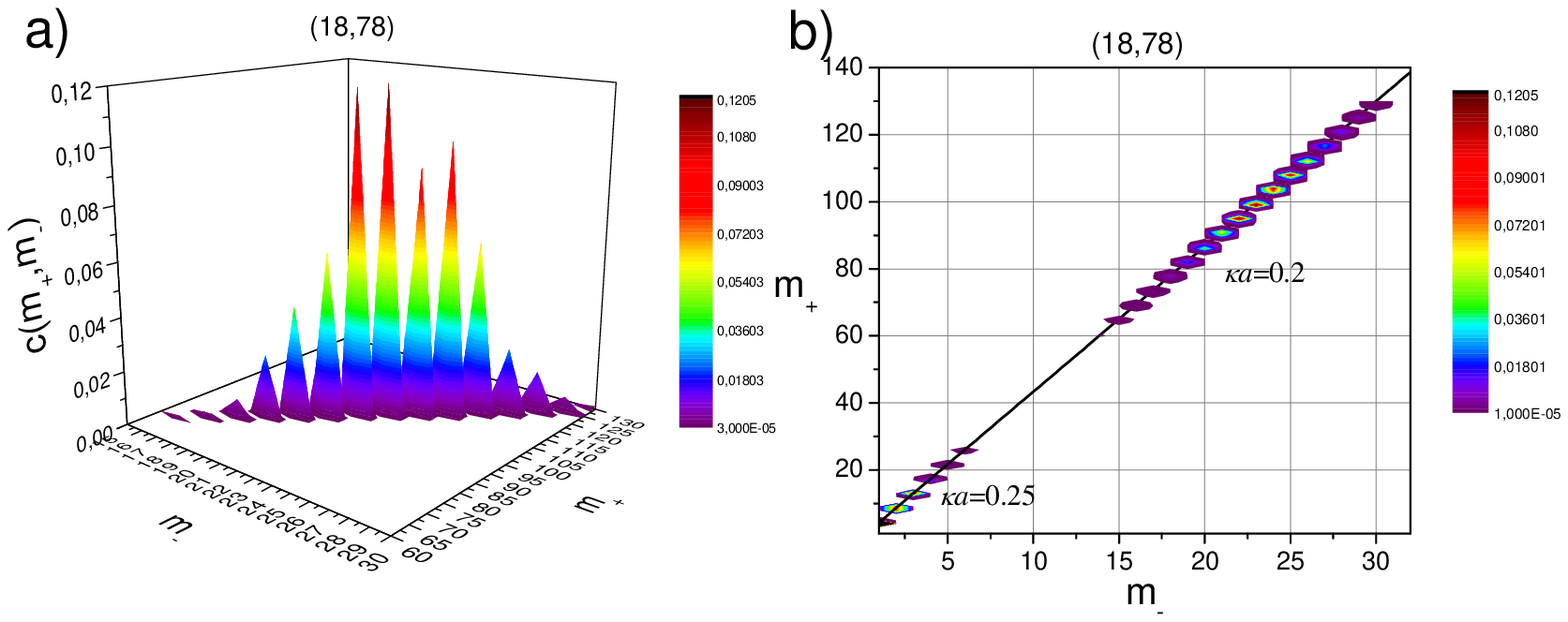}
\end{center}
\caption{Case A. a) Probability distribution function (normalized
$vc_{m_{+},m_{-}}$) for asymmetric block copolymers of opposite
charge, $z_+=18$ and $z_-=78$ with the same distance between the
charges, and Manning parameter $\xi=0.25$. The length of neutral
block (stabilizing corona) is $N=200$, the salt concentration is
$\kappa a=0.2$, and the concentrations of individual chains is
$vc_{1,0}=10^{-4}$ and $vc_{0,1}=10^{-9}$. b) Projection of the
same function on the plane $(m_+,m_-)$ for three different salt
concentrations.} \label{Fig:2D}
\end{figure*}

\subsection{Results}
Eq. (\ref{eq:cmp}) defines the equilibrium distribution function
of the complexes $c_{m_{+},m_{-}}$ as a function of the geometry
of the chains, asymmetry of the charges along the chain and salt
concentration. We assume that the conformation of the
polyelectrolyte complex is a spherical aggregate with a core
formed by charged blocks surrounded by neutral corona (Figure
\ref{Fig:Polymic}-A)). Since the electrostatic contribution of the
core and isolated chains in the solution is the main contribution
to the free energy, one might expect that the thermodynamically
stable complexes would be narrowly distributed in size and have
the minimal possible charge. Thus, the equilibrium of the free
energy would require the compensation of the charges inside the
core, such that the formed PIC micelles are almost neutral.
However, in our description we allow for deviations from zero
charge, because other contributions to the total free energy, the
entropy of mixing, the salt concentration and the steric repulsion
in the corona, may shift the equilibrium.

\begin{figure}
\begin{center}
\includegraphics[width=8.25cm,clip]{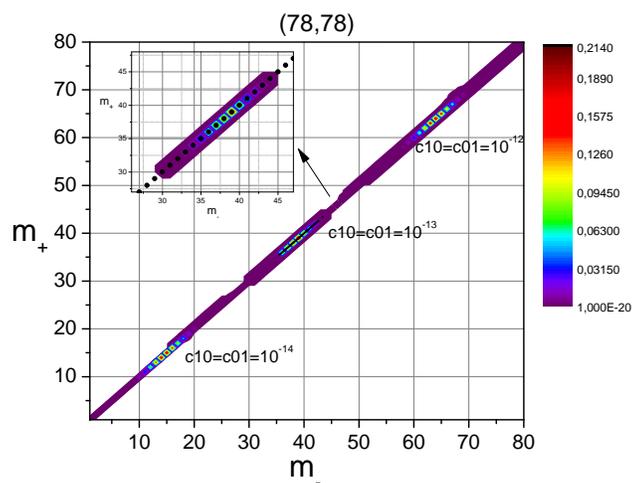}
\end{center}
\caption{Projections of the probability distribution function
(normalized $vc_{m_{+},m_{-}}$) on the plane ($m_{+},m_{-}$) for
the equally charged copolymers of opposite charges, $z_+=78$ and $z_-=78$
as a function of
polymer concentrations, $vc_{1,0}$ and $vc_{0,1}$. Manning parameter,
$\xi=0.25$
and salt concentration $\kappa a=0.2$. The inset shows the points of
electroneutrality of the complexes (black disks) together
with a zoom onto the  $vc_{1,0}=vc_{0,1}=10^{-12}$ case.} \label{Fig:conc}
\end{figure}

The distribution function of the complexes, Eq. (\ref{eq:cmp}) is
calculated for each combination of ($m_+$, $m_-$) and the results
are shown in Fig. \ref{Fig:2D}a). As an example, we plot the
normalized $ vc_{m_{+},m_{-}}$ for a mixture of a linear polymer
with the charge $z_{+}=18$ and oppositely charged diblock
copolymer with a charged block, $z_{-}=78$, and neutral block of
length $N=200$ (case A). The two polymers share the same distance
between the charges along the chain: $n_{+}=n_{-}=1/\xi=4$. It can
be noted that all three distributions reported lie around the
``electroneutrality line'' $z_{+}m_{+}=z_{-}m_{-}$.
In the vicinity of that line, the precise location of the support
of the distribution function stems from a subtle balance of
effects, as embodied in the free energy (\ref{eq:free}). We
observe in Fig. \ref{Fig:2D}b) that upon increasing the salt
concentration, the equilibrium size distribution is shifted
towards smaller complex sizes and becomes more peaked. For
instance, for $\kappa a =0.3$, the peak corresponds to $m_+=24$
and $m_-=7$ (for which the complex has charge
$z_{+}m_{+}-z_{-}m_{-}=-72$). A similar trend is observed while
changing the length of the neutral block, which controls the
repulsion in the corona. Long tails in the corona favor smaller
complexes, and shift the equilibrium accordingly. Since the
complexes are close to neutrality, the salt concentration mostly
affects the electrostatic energy of free chains [Eq.
(\ref{OmEl})], and the shift of the aggregation numbers along the
electroneutrality line is mainly due to the chains in the
solution. In addition, increasing the bulk concentration of
polymers, $vc_{1,0}$ and $vc_{0,1}$, increases the aggregation
numbers, see Fig. \ref{Fig:conc}.

\begin{figure}
\begin{center}
\includegraphics[width=8.25cm,clip]{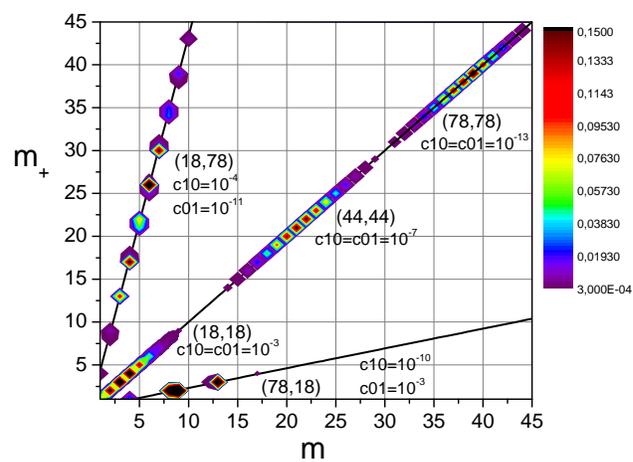}
\end{center}
\caption{Projections of the probability distribution function
(normalized $vc_{m_{+},m_{-}}$) on the plane ($m_{+},m_{-}$) for
different lengths of block copolymers of opposite charge with the
same distance between the charges,  $\xi=0.25$
and $\kappa a=0.2$. Increasing concentrations
of individual chains, $vc_{1,0}$ and $vc_{0,1}$, increases the
aggregation numbers, which move along the electroneutrality
lines.} \label{Fig:2DHarada}
\end{figure}

Fig. \ref{Fig:2DHarada} shows the size distribution function of
the complexes formed by equally charged ("matched" in terms of
Ref. \citenum{Harada}) polymers (18,18), (44,44) and (78,78) and
"unmatched" polymers, (18,78) and (78,18). The polymer concentrations
are chosen in such a way that the complexes are
formed close to the origin, which may indicate the onset of
aggregation.
Aggregation of long polymers, (78,78), occurs at smaller
concentrations than aggregation of short polymers, (18,18), due to
the entropy of mixing, which strongly depends on the total length
of polymers. We find that unmatched complexes [see the cases
(18,78) and (78,18)], can also be formed if the aggregation
numbers are close to the electroneutrality line (the opposite
charges are compensated). On the other hand, Ref \cite{Harada} has
put forward a chain recognition mechanism where matched cases are
more prone to form large complexes, but the system considered
there is somewhat different, involving the equilibrium between
three types of individual chains together with two and three
component complexes.


\section{Discussion and conclusions}
\label{sec:concl}

We have developed a framework to study the formation of
polyelectrolyte complexes from an initial arbitrary mixture of
charged polymers, where both polycations and polyanions are
present in an electrolyte solution. Two situations were addressed,
as sketched in Fig. \ref{Fig:Polymic}: for a given polycation
type, the polyanion is either a diblock copolymer with a long
neutral tail (case A), or a polyanion having a different
intercharge spacing along the backbone (case B). Coulombic
attraction between oppositely charged polymers leads to the
formation of complexes, with an {\it a priori} unknown
composition. The numbers of chains of both types in a given
complex were denoted $m_+$ and $m_-$. These complexes were
envisioned as forming hairy structures, where the hair/corona is
either made up of dangling neutral chains (case A) or of loops
(case B), while the core of much smaller spatial extension
contains most of the charges of the polymeric backbones. We
started by focusing on the electrostatic aspects, treated at the
level of Poisson-Boltzmann theory. This part of the work thereby
extends a previous study performed for salt-free systems
\cite{Raphael}. In a second step, the resulting electrostatic free
energy of the complexes was used, together with the entropic
repulsion between tails/loops in the corona, to provide us with a
free energy functional for an arbitrary mixture of complexes
having a given size distribution $c_{m_+,m_-}$. Upon minimizing
this functional under the appropriate constraints of mass
conservation for both polycationic and polyanionic species, we
obtained the equilibrium composition of our mixture. Whereas this
optimal distribution turns out to give a negligible weight to
configurations that depart from complex global charge neutrality
--a property that may have been anticipated--, it exhibits the
non- trivial feature of a high selectivity: out of an initial
random soup of polycations and polyanions, well defined complexes
with precise composition ($m_+$, $m_-$) may emerge, particularly
when the salt density is increased.

The problem under study here is characterized by a large number of
dimensionless parameters, and we furthermore made simplifying
assumptions in the description, such as equating the Kuhn lengths
for both positively and negatively charged polymers. We chiefly
focused on the effect of changing the salt concentration, which is
an experimentally simple control parameter. The pH dependence of
the core charge of the complexes has not been addressed, but it
can be incorporated for instance via the Henderson--Hasselbalch
equation \cite{Semenov}. In addition, the Coulombic aspects were
treated at mean-field level, which is adequate provided the bare
charge of the complex core, $Z_1$, is smaller than a bound of
order $[R_c/(z^2 \ell_B)]^3$ \cite{Chepelianskii}, which decreases
when increasing the valence $z$ of the mobile micro-ions (assumed
here monovalent, i.e. $z=1$). Finally, we have neglected the
structure of the core, by homogeneously smearing out its charge.
This certainly leads to overestimate their free energy, due to the
neglect of the corresponding negative correlation energy
\cite{Levin}.


\section*{Summary of main notations used}

\begin{table}[ht]
\scriptsize
\begin{tabular*}{0.5\textwidth}{@{\extracolsep{\fill}}lll}

$q$ & & elementary charge \\
$a$ & & Kuhn length, assumed equal for both polycationic and polyanionic
chains\\
$z_{\pm}$  & & total charge of a chain, in units of  $\pm q$ \\
$n_\pm$    & & distance between $\pm q$ charges along a linear polymer,
                     in units of Kuhn length \\
$N$        & & length of a neutral polymer block, in units of the Kuhn
length \\
$m_\pm$    & & number of positive/negative chains in an aggregate
(core)\\
$\ell_B$   & & Bjerrum length $q^2/(\epsilon kT)$ defined from
temperature
and solvent dielectric \\
&&permittivity\\
$R_c$      & & radius of a spherical aggregate/core\\
$Z_1$      & & bare charge of a spherical core (due to polymers)\\
$Z_2$      & & ``uptake'' charge of a core (due to polymers and salt ions
                     inside the core)\\
$Z_3$      & & effective (or renormalized) charge of a spherical core,
relevant at large\\
&&   distances from the core center\\
$\widetilde Z$ & & \hbox{reduced charge, defined as $\widetilde Z = Z
\ell_B
/R_c$}\\
$c_{\infty}$ & & salt density in the reservoir\\
$\kappa^{-1}$ && Debye length, defined through $\kappa^2 = 8 \pi \ell_B
c_{\infty}$      \\
$\rho(r)$    & & total density of charge at a distance $r$ from core
center\\
$\rho_c$    & & parameter controlling the charge of the corona (case B)\\
$\xi$       & & Manning parameter defined as $\lambda_{\pm} \ell_B
\propto
\ell_B/n_{\pm}$, where $\lambda_\pm$ is the linear charge\\
&&of a linear polymer\\

\end{tabular*}
\end{table}

\section*{Acknowledgments}
We would like to thank A. Chepelianskii, F. Closa and E. Raphael
for useful discussions.

\footnotesize{

\providecommand*{\mcitethebibliography}{\thebibliography}
\csname @ifundefined\endcsname{endmcitethebibliography}
{\let\endmcitethebibliography\endthebibliography}{}

}

\balance

\end{document}